\documentclass[aps,pra,twocolumn,superscriptaddress,groupedaddress,nofootinbib]{revtex4-1}  
\usepackage{graphicx}  
\usepackage{dcolumn}   
\usepackage{bm}        
\usepackage{amssymb}   
\graphicspath{{figures/}} 
\usepackage{epstopdf}	
\usepackage{CJK}
\usepackage{amsmath} 

\usepackage{textcomp}

\begin{document}
	
	\begin{CJK*}{}{} 
		\title{Demonstration of Ramsey-Comb Precision Spectroscopy in Xenon at Vacuum Ultraviolet Wavelengths Produced with High-Harmonic Generation}
		\author{L.S.~Dreissen}
		\author{C.~Roth}
		\author{E.L.~Gr\"{u}ndeman}
		\author{J.J.~Krauth}
		\author{M.G.J.~Favier}
		\author{K.S.E.~Eikema}
		\affiliation{LaserLaB, Department of Physics and Astronomy, Vrije Universiteit, De Boelelaan 1081, 1081 HV Amsterdam, The Netherlands}
		\date{\today}
		
		\begin{abstract}
			The remarkable progress in the field of laser spectroscopy induced by the invention of the frequency-comb laser has enabled many new high-precision tests of fundamental theory and searches for new physics. Extending frequency-comb based spectroscopy techniques to the vacuum and extreme ultraviolet spectral range would enable frequency measurements of transitions in e.g.~heavier hydrogen-like systems and open up new possibilities for tests of quantum electrodynamics and measurements of fundamental constants. The two main approaches,  full-repetition rate up-conversion in a resonator, and two-pulse amplification and up-conversion for the Ramsey-comb technique, rely on high-harmonic generation (HHG), which is known to induce spurious phase shifts from plasma formation. After our initial report~\cite{dreissen_high-precision_2019}, in this article we give a detailed account of how the Ramsey-comb spectroscopy technique is used to probe the dynamics of this plasma with high precision, and enables accurate spectroscopy in the vacuum ultraviolet. It is based on recording Ramsey fringes that track the phase evolution of a superposition state in xenon atoms excited by two up-converted frequency-comb laser pulses. In this manner, phase shifts up to 1 rad induced by the HHG process could be observed at nanosecond timescales with mrad-level accuracy at 110 nm. We also  show that such phase shifts can be reduced to a negligible level of a few mrad. As a result we were able to measure the $5p^6 \rightarrow 5p^5 8s~^2[3/2]_1$ transition in $^{132}$Xe at 110 nm (the seventh harmonic of 770 nm) with sub-MHz accuracy, leading to a transition frequency of $2\,726\,086\,012\,471(630)$ kHz. This value is 10$^4$ times more precise than the previous determination and the fractional accuracy of $2.3 \times 10^{-10}$ is 3.6 times better than the previous best spectroscopic measurement using high-harmonic generation. Additionally, the isotope shifts between $^{132}$Xe and two other isotopes ($^{134}$Xe and $^{136}$Xe) were determined with an accuracy of 420 kHz. The method can be readily extended to achieve kHz-level accuracy by increasing the pulse delay, e.g.~to measure the $1S-2S$ transition in He$^+$. Therefore, the Ramsey-comb method shows great promise for high-precision spectroscopy of targets requiring vacuum ultraviolet and extreme ultraviolet wavelengths. 
		\end{abstract}
		
		\maketitle
	\end{CJK*}
	
	\section{Introduction}
	\label{sec:intro}
	
	Precision spectroscopy in calculable systems enables tests of fundamental theory and searches for physics beyond the standard model. Recent experimental advances have led to unprecedented accuracy and the most stringent test of bound-state quantum electrodynamics theory (QED)~\cite{parthey_improved_2011,holsch_benchmarking_2019,biesheuvel_probing_2016,stohlker_1_2000,kozlov_highly_2018,draginic_high_2003}. Further improvements are hampered by the uncertainty of experimental parameters such as the nuclear charge radius, which influences the energy structure through finite nuclear size effects. This problem can be inverted to extract improved values for these parameters using spectroscopic measurements in combination with accurate theoretical calculations. In 2010, this instigated the so-called \textit{proton radius puzzle}~\cite{pohl_size_2010}, where the inferred proton radius from spectroscopy on muonic hydrogen was 4\% smaller than that obtained from electronic hydrogen spectroscopy~\cite{parthey_improved_2011,mohr_codata_2016}, which was equal to a 4.4$\sigma$ discrepancy~\cite{pohl_size_2010,antognini_proton_2013,pohl_muonic_2013,antognini_experiments_2016,pohl_laser_2016,mohr_codata_2016}. However, three out of four recent measurements have now confirmed the smaller (muonic) proton radius~\cite{bezginov_measurement_2019,beyer_rydberg_2017,fleurbaey_new_2018}, including the most recent result from electron-proton scattering~\cite{xiong_small_2019}. This makes new targets with a bigger nuclear-size induced frequency shift, such as the He$^+$ ion, particularly interesting~\cite{krauth_paving_2019}, as similar comparisons with muonic He$^+$ can be made~\cite{franke_theory_2017,diepold_theory_2018}. Especially because the QED contributions are much larger in He$^+$ than in hydrogen as a result of the higher-order scaling with the nuclear charge $Z$ of these terms. Therefore, in combination with the expected improved determination of the alpha radius from muonic He$^+$~\cite{franke_theory_2017,diepold_theory_2018}, this can lead to a more stringent test of QED. With further theoretical improvements, also a value of the Rydberg constant can be determined from He$^+$ spectroscopy, which is effectively independent from hydrogen spectroscopy (because the required alpha particle charge radius, determined from muonic He$^+$ spectroscopy, only weakly depends on the Rydberg constant). The obvious challenge for a measurement of the $1S-2S$ transition in He$^+$ arises from the short excitation wavelength, which lies in the extreme ultraviolet spectral range~\cite{herrmann_feasibility_2009}.\\
	Spectroscopic measurements at these short wavelengths can be realized by using e.g.~radiation from a large synchrotron facility to perform Fourier-transform spectroscopy at wavelengths down to 40 nm~\cite{oliveira_fourier_2009,oliveira_high-resolution_2011}. In this manner, relative accuracies of $1\times 10^{-7}$ have been demonstrated~\cite{heays_high-resolution_2011}. However, a higher accuracy can be reached with laser spectroscopy in combination with nonlinear up-conversion. For example, amplified and up-converted nanosecond pulses have been used to reach a fractional accuracy of $0.9\times 10^{-8}$ at 58 nm~\cite{eikema_precision_1996}. This approach is based on traditional  frequency-scanning spectroscopy, but uncertainties due to frequency-chirp induced by the amplification process and the relatively broad bandwidth of the excitation pulses ultimately limit this approach. A much higher spectroscopic accuracy can be reached with pulses from a frequency-comb (FC) laser \cite{holzwarth_optical_2000,jones_carrier-envelope_2000}.\\ 
	The pulses emitted by a FC laser have a typical pulse energy of a few nJ, which is insufficient for up-conversion to the vacuum ultraviolet (VUV) and extreme ultraviolet (XUV) spectral range using high-harmonic generation. Therefore the peak intensity of these pulses has to be increased to reach this part of the spectrum. One approach is based on intra-cavity HHG to up-convert the full frequency-comb pulse train to the XUV spectral range ~\cite{jones_phase-coherent_2005,gohle_frequency_2005}. In 2005 this technique was first demonstrated and in 2012 Cing\"{o}z \textit{et al.}~have shown that they could perform spectroscopy in argon at 82 nm with 3 MHz accuracy (0.8 ppb) using such an XUV-comb~\cite{cingoz_direct_2012}. \\
	Alternatively, a single pair of FC pulses can be amplified to the mJ-level using chirped-pulse amplification followed by subsequent up-conversion using HHG in a single pass. In 2010 Kandula \textit{et al.}~demonstrated this technique and performed a Ramsey-type experiment in neutral helium atoms where they achieved an accuracy of 6 MHz at 51 nm (1 ppb) \cite{kandula_extreme_2010}. This accuracy was limited by detrimental phase shifts arising from amplification and up-conversion. \\
	We use a modified version of the latter spectroscopy method, called the Ramsey-comb spectroscopy (RCS) technique~\cite{morgenweg_ramsey-comb_2014}, that bypasses these limitations. It is based on multiple recordings with pairs of up-converted FC pulses to obtain a \textit{series} of Ramsey fringes from which the transition frequency can be extracted by comparing the relative phase. As a result, common mode phase shifts arising from e.g.~amplification and up-conversion, but also the ac-Stark shift, are suppressed strongly~\cite{morgenweg_ramsey-comb_theory_2014,altmann_deep-ultraviolet_2018}.\\ 
	The technique combines the high precision from a FC laser with the advantages of pulse amplification, leading to e.g.~easy tunability, simple up-conversion in a gas jet (no resonators required), high excitation probability and a nearly 100$\%$ detection efficiency, because the detection can be carefully timed (e.g.~using a state-selective ionization pulse). The method has been demonstrated to work very well over a broad spectral range (from NIR to DUV)~\cite{morgenweg_ramsey-comb_2014,altmann_high-precision_2016,altmann_deep-ultraviolet_2018}. We recently showed that RCS can be combined with HHG to extend the method to the VUV and XUV spectral range~\cite{dreissen_high-precision_2019}. The extension of RCS with HHG was so far hampered by the unknown detrimental phase shifts which arise from ionization during up-conversion \cite{corsi_direct_2006,bohan_phase-dependent_1998,brandi_spectral_2006}. The influence from plasma formation has been previously investigated only on short timescales, ranging from a few hundred femto-seconds to a few picoseconds~\cite{salieres_frequency-domain_1999,chen_measurement_2007}, but not on timescales relevant to RCS (i.e.~at tens to hundreds of nanoseconds).\\
	In this paper we give a detailed account of how the RCS method can be used to provide insights into plasma-induced phase shifts arising from HHG by tracking the atomic phase evolution of excited atoms (xenon) with mrad-scale precision. Moreover, we show how this results in high-precision spectroscopy with light from HHG in the vacuum ultraviolet on the $5p^{6} \rightarrow 5p^{5}8s [3/2]_1$ transition in xenon at 110 nm.
	
	\section{Experimental techniques}
	\label{sec:exp_princples}
	In the original Ramsey spectroscopy scheme, molecules or atoms are excited by two separated phase-locked oscillatory fields in the radio-frequency (RF) domain~\cite{ramsey_molecular_1950}. The Ramsey-comb spectroscopy (RCS) technique~\cite{morgenweg_ramsey-comb_2014} also uses two interaction pulses, separated in time instead of space and in the optical domain, based on pairs of phase-locked laser pulses from a frequency-comb laser.
	
	\subsection{The Ramsey-comb spectroscopy technique}
	\label{subsec:RCS}
	A two-level system with transition frequency $f_{tr}$ can be brought in a superposition of the ground and the excited state by a resonant laser pulse. Excitation with two phase-locked pulses leads to interference between two such contributions. The interference pattern can be probed by observing the excited state population (e.g.~through detection of fluorescence, or more efficiently, by photo-ionization). The phase evolution of the interfering contributions depends on the delay ($\Delta t$) and the phase difference ($\Delta \phi$) between the pulses, leading to an oscillating upper state population ($|c_{e}(\Delta t, \Delta \phi)|^2$), which is described by
	\begin{equation}
	|c_{e}(\Delta t, \Delta \phi)|^2 \propto \cos{(2 \pi f_{tr} \Delta t + \Delta \phi)}.
	\end{equation}
	A FC laser is a very suitable source for this excitation scheme, because it provides accurate control over the pulse delay through the pulse repetition time ($T_{rep}$) and the relative phase through the carrier-envelope phase slip ($\Delta \phi_{ceo}$). One can adjust either one of these parameters to obtain a Ramsey fringe, but experimentally it is more convenient to adjust the delay $\Delta t = T_{rep} + \delta t$. The range over which the delay is scanned, is related to the transition frequency and is typically on the order of a few hundred attoseconds.\\
	The amplification process induces a phase shift of up to 300 mrad on the excitation pulses ($\Delta \phi_{opt}$), which is added to the carrier-envelope offset phase $\Delta \phi = \Delta \phi_{ceo}+\Delta \phi_{opt}$. This contribution is typically hard to determine with high accuracy. However, it is possible to construct an amplifier based on parametric amplification that can keep the phase shift between the two amplified pulses constant at the mrad-level as a function of delay.
	\begin{figure}[!t]
		\centering
		\includegraphics{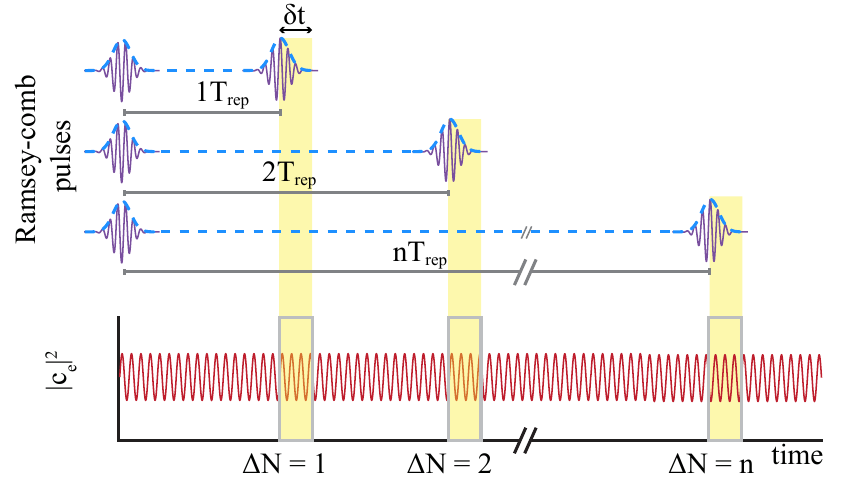}
		\caption{A series of Ramsey fringes from two broadband resonant FC pulses is recorded at different multiples of the repetition time ($\Delta N \times T_{rep}$). Each individual fringe is obtained by changing the delay by $\delta t$, which is on an attosecond timescale. The transition frequency is extracted from the relative phases of the fringes, suppressing the influence of common mode phase shifts of the Ramsey fringes on the extracted transition frequency.}
		\label{fig:RCS_explained}
	\end{figure}
	Therefore in RCS one records a series of Ramsey fringes at different pulse delays separated by an integer times the repetition time ($\Delta N \times T_{rep}$), as illustrated in Fig.~\ref{fig:RCS_explained}. The transition frequency is obtained by comparing the relative phase of these fringes, leading to a suppression of common mode shifts. A constant phase shift between the amplified FC pulses manifests itself as a global phase shift of the fringes and does not influence the extracted transition frequency. This argument also holds for other common mode effects, such as the ac-Stark shift~\cite{morgenweg_ramsey-comb_theory_2014,altmann_deep-ultraviolet_2018}. Additionally, the accuracy of the measurement improves proportionally with the inter-pulse delay, as the frequency that fits the measured phase evolution as a function of the delay time becomes more sensitive for longer delays.\\
	The transition frequency extracted in RCS is affected by both the first-order Doppler shift and the recoil shift from the absorption of a single photon during excitation. Cadoret \textit{et al.}~have shown that the phase $\Phi$ of the superposition of ground and excited state after a free evolution time $\Delta t$ in a Ramsey interferometer can be written as \cite{cadoret_atom_2009} 
	\begin{equation}
	\label{eq:recoil_phase}
	\Phi = \omega \Delta t +\Delta \phi+ \Big(\frac{\mathbf{v}_e+\mathbf{v}_g}{2}\Big)\cdot(\mathbf{v}_e-\mathbf{v}_g)\frac{m \Delta t}{\hbar},
	\end{equation}
	where $\omega$ is the excitation frequency, $\mathbf{v}_g$ and $\mathbf{v}_e$ are the velocities of the atoms in the ground and excited state, respectively, and $m$ is the mass of the atom. The recoil velocity after excitation is given by $\mathbf{v}_e-\mathbf{v}_g = \hbar \mathbf{k}/m$. Substituting this into Eq.~\ref{eq:recoil_phase} and taking only the velocity component along the propagation direction $z$ of the excitation laser into account, one obtains
	\begin{equation}
	\label{eq:recoil_phase_2}
	\Phi = \omega \Delta T +\Delta \phi+ (v_{e,z}+v_{g,z})\frac{\omega \Delta t}{2c}, 
	\end{equation}
	where $c$ is the speed of light. The last term in Eq.~\ref{eq:recoil_phase_2} is a combination of Doppler and recoil shift, which can be viewed as a generalized Doppler shift~\cite{barnett_recoil_2010}. It also means that in Ramsey-comb spectroscopy the recoil frequency shift has to be taken into account for one-photon excitation.\\
	Note that in two-photon spectroscopy in a counter-propagating configuration ($k_1 = -k_2$) \cite{altmann_high-precision_2016,altmann_deep-ultraviolet_2018}, the recoil shift is canceled and does not influence the extracted transition frequency.
	
	\subsection{Combining RCS with HHG}
	\label{subsec:RCS_HHG}
\begin{figure*}[!t]
		\centering
		\includegraphics{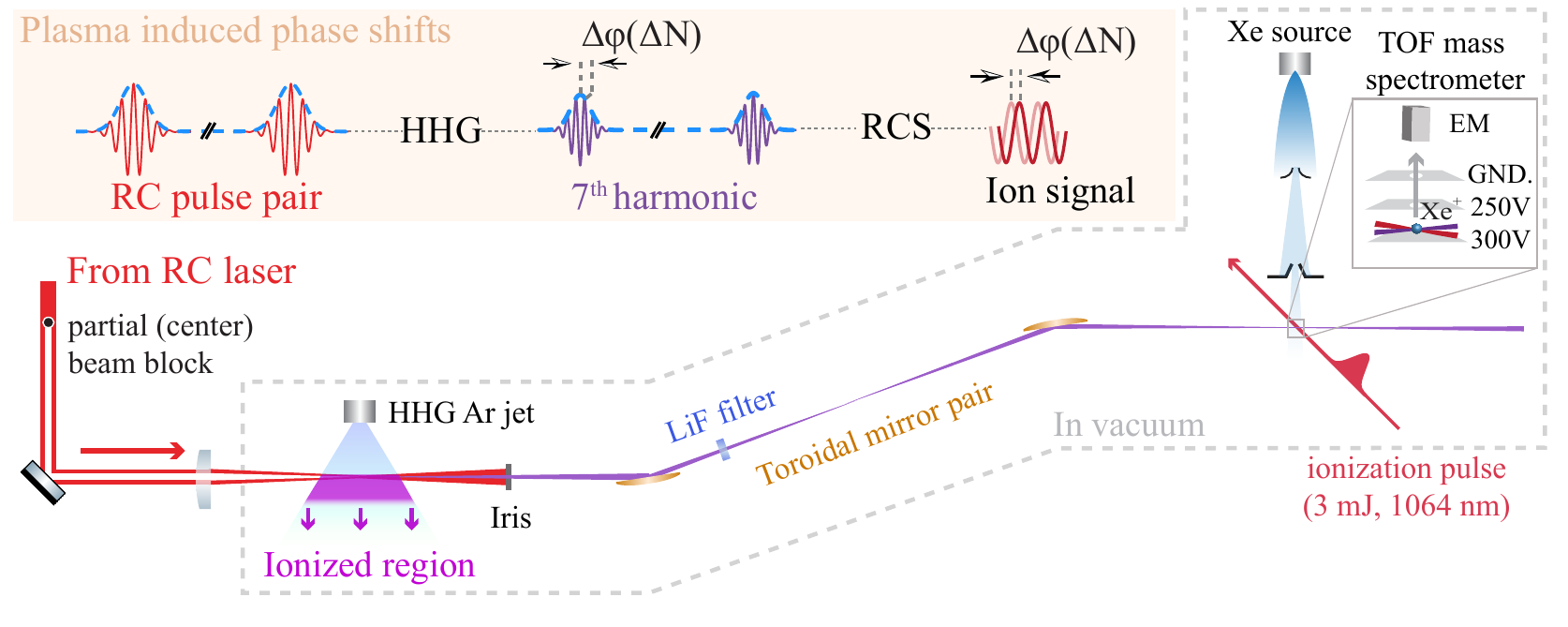}
		\caption{The experimental setup and a schematic representation of the influence from plasma formation during HHG on RCS (top panel). The NIR beam from the RC laser is focused to generate high-harmonics in argon. A combination of a beam block and an adjustable iris is used to separate the harmonics from the fundamental beam, while a LiF plate acts as a filter that only transmits light $\lambda > 105$ nm. The VUV beam is refocused using a toroidal mirror pair at grazing incidence and overlapped with a pulsed xenon beam at 90$^{\circ}$ angle. An ionization pulse at 1064 nm is delayed by 2 ns with respect to the second VUV pulse and selectively ionizes the excited atoms. The ions are extracted with a pulsed electric field and detected through a time-of-flight (TOF) drift tube by an ETP AF880  electron multiplier (EM). The plasma induced delay-dependent phase shift $\Delta \phi (\Delta N)$ between the two RC pulses from HHG, is detected through the phase of the observed ion signal Ramsey fringes (top panel).}
		\label{fig:vacuum_setup}
	\end{figure*}

	Higher-order harmonic generation uses the nonlinear response of a gas to coherently up-convert frequencies from the infrared or visible spectral range to the vacuum ultraviolet or extreme ultraviolet spectral range.
	An intuitive picture of HHG is given at a single-atom level by the semi-classical three-step recollision model~\cite{corkum_plasma_1993} (a quantum-mechanical description is given by the Lewenstein model~\cite{lewenstein_theory_1994}).\\ 
	The principle is based on focusing a powerful short laser pulse in a medium (typically a gas jet, or a gas filled capillary) to reach an intensity of approximately $10^{14}$ W/cm$^2$. At these intensities, the electric field of the laser pulse can significantly perturb the Coulomb potential of the least bound electrons, and for a high enough field this enables the electron to tunnel out into the continuum. After ionization, the electron gains energy as it is accelerated in the strong electric field of the laser. It can return to the parent ion because the field of the fundamental laser changes sign after half a period of the optical wave. When it returns it can recombine, leading to the emission of a high-energy photon. The recollision probability is highest near every zero-crossing of the electric field and the process is therefore tightly linked in time to the fundamental wave, leading to a high degree of coherence. This has been demonstrated in several experiments involving interference of HHG beams, see e.g.~\cite{zerne_phase-locked_1997, bellini_temporal_1998}, and by the high coherence of steady-state full-repetition rate frequency comb up-conversion~\cite{benko_extreme_2014}. For a symmetric medium, such as a simple gas jet, only odd harmonics of the fundamental optical carrier frequency are produced. The efficiency of the HHG process is often very low (typically $10^{-6}$ in a simple gas jet) and most electrons might not recombine but instead become permanently ionized. Over time a significant fraction of the medium can become ionized and a plasma can be formed. In order to combine RCS with HHG, the effect of this plasma formation on the phase of the up-converted FC pulses has to be carefully considered.\\
	In RCS common mode effects are significantly suppressed and therefore the plasma build-up during each pulse does not influence the extracted frequency. If the pulses do not influence each other, the plasma-induced shift is equal for both up-converted FC pulses. However, the second pulse can experience a phase shift due to a change in the refractive index of the HHG medium because of the plasma formed during the first pulse. This shift is dependent on the inter-pulse delay because the ions and electrons move out of the interaction region, and partly recombine. It therefore leads to a delay-dependent phase shift and, as a result, a shift of the extracted transition frequency. On the other hand, because this delay-dependent plasma-induced phase shift is directly imprinted on the phase of the Ramsey fringe, as is illustrated in the top panel of Fig.~\ref{fig:vacuum_setup}, it can be detected by tracking the phase evolution of the atomic superposition state at different inter-pulse delay $\Delta N$ and driving intensity. The results presented in Sec.~\ref{subsec:RCS_HHG} rely on this principle.\\
	The magnitude of the induced shift is determined by three contributions to the change of the refractive index: the free electrons, the generated ions and the depletion of neutral atom density~\cite{rundquist_phase-matched_1998,popmintchev_phase_2009}. The influence from the free electrons in the plasma is dominant over the contribution from the atom depletion and the ions. The latter is usually neglected, because the energy levels scale up for higher charge states, leading to a refractive index close to unity. The different contributions will be discussed in more detail in Sec.~\ref{subsec:plasma_induced_effects}.\\
	Another effect, which has to be considered when combining HHG with RCS, is the influence of phase noise. In HHG the amplitude of the phase noise originating from the frequency comb and the amplification process is multiplied-up by the harmonic order, and can therefore lead to a significant reduction of contrast of the Ramsey fringes. The phase noise amplitude should remain well below $\pi$ at the excitation wavelength to retain enough contrast ($>20\%$) for a RCS measurement.
	\begin{figure}[!t]
		\centering
		\includegraphics{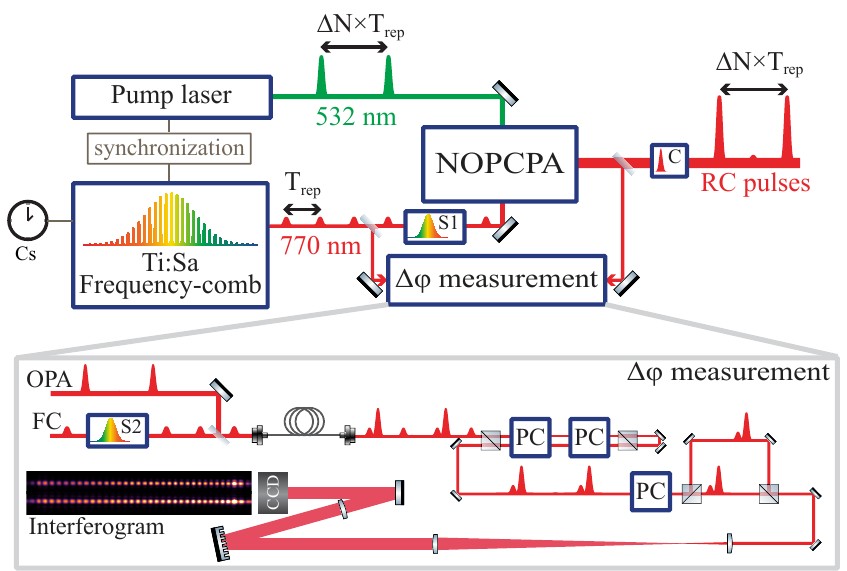}
		\caption{Pulse pairs with a delay of $\Delta N \times T_{rep}$ from a Ti:sapphire frequency-comb laser are stretched ($S1$) and selectively amplified in a non-colinear optical parametric chirped pulse amplifier (NOPCPA). The pulse pair is selected by adjusting the settings of the pump laser. A typical energy of 3 mJ/pulse is reached of which the main part is compressed in a grating compressor ($C$) and up-converted using high-harmonic generation. A small fraction of the amplified beam is used to measure the phase difference between the pulses in the setup shown in the lower panel. For this purpose a second stretcher ($S2$) is used to stretch the reference FC pulses after which they are spatially overlapped with the amplified pulses in a single mode fiber. The pulse pairs are separated from the full FC pulse train using two Pockels cells ($PC$) in double pass. A third PC is used to introduce a slight vertical displacement for the first or the second pulse pair in order to project them separately on the CCD camera. By introducing a small time delay (typically $<1$ ps) between the amplified and reference pulses, a spectral interferogram is recorded using a spectrometer based on a grating and a CCD camera from which the phase difference between the amplified pulses is obtained.}
		\label{fig:RC_laser_setup}
	\end{figure}

	\section{Experimental setup}
	\label{sec:experiment}
	The RCS setup is based on selective amplification of a FC pulse pair using parametric chirped pulse amplification. The re-compressed amplified pulses have a sufficiently high peak-intensity to create high-harmonics in a simple single-pass gas jet. After up-conversion, the pulse pair is refocussed in an atomic xenon beam to excite the $5p^{6} \rightarrow 5p^{5}8s [3/2]_1$ transition at 110 nm. The excited atoms are state-selectively ionized and detected using a time-of-flight mass spectrometer. In this section, the individual components of the system are described in detail.
	\subsection{The Ramsey-comb laser system}
	\label{subsec:RCS_laser}	
	An overview of the laser setup is shown in Fig.~\ref{fig:RC_laser_setup}. The starting point is a Kerr-lens mode-locked titanium:sapphire frequency-comb laser with an average output power of 450 mW and a repetition frequency of $f_{rep}$ = 126.6 MHz. The spectrum of the laser is centered around 800 nm and has a bandwidth of $\sim$75 nm. The repetition frequency and the carrier-envelope offset frequency ($f_{ceo}$) are stabilized using a commercial cesium atomic clock (Symmetricon CsIII 4310B). This enables accurate control over the repetition time $T_{rep}=1/f_{rep}$ and the carrier-envelope phase slip $\Delta \phi_{ceo} = 2\pi f_{ceo} / f_{rep}$.\\
	The FC pulses are stretched and spectrally clipped in a 4f-grating stretcher to improve the temporal overlap of the FC pulses with the pump pulses in the parametric amplifier. The stretcher introduces a group velocity dispersion (GVD) of $1.5 \times10^{6}$ fs$^{2}$ and an adjustable slit in the Fourier plane of the stretcher selects a spectrum with a bandwidth of typically 8 nm that is centered around 770 nm. The combination of stretching and spectral clipping leads to a pulse length of $\sim$10 ps.\\
	A three stage non-collinear optical parametric chirped pulse amplifier (NOPCPA) based on three beta-barium borate (BBO) crystals is then used to amplify the FC pulses. It is pumped by a home-built laser system providing a pulse pair at 532 nm with an energy of 17 mJ/pulse and a pulse length (full-width at half maximum) of 60 ps~\cite{morgenweg_multi-delay_2013,morgenweg_1.8_2012}. The inter-pulse delay between the pump pulses can be adjusted by an integer number $\Delta N$ of the repetition time ($T_{rep}$) with a set of modulators \cite{morgenweg_multi-delay_2013}, leading to selective amplification of the corresponding FC pulse pair. A typical energy of 3 mJ/pulse is reached in this manner.\\
	The amplified pulses are re-compressed to a pulse length of 220 fs (full-width at half maximum) using a grating compressor, based on transmission gratings (Lightsmyth T-1850-800s). One of the gratings in the compressor is mounted on a translation stage so that the amount of induced GVD can be adjusted. The whole compressor is build on a rotation stage to tune the angle of incidence and compensate for third-order dispersion. However, the hard edge used for wavelength selection in the stretcher leads to a sinc-like pulse shape and therefore to pre- and after-pulses of a few percent with respect to the main peak. The remaining pulse energy after re-compression is typically 2 mJ/pulse.
	\subsection{The phase-measurement setup}
	\label{subsec:phase_measurement_setup}
	The phase difference between the amplified FC pulses is measured in a separate setup, shown in the lower panel of Fig.~\ref{fig:RC_laser_setup}, which is based on spectral interference. For this purpose a small fraction of the amplified beam is split-off before the compressor and spatially overlapped with the original unamplified FC pulse in a single-mode fiber. Saturation effects in the parametric amplifier lead to a slightly wider spectrum of the amplified pulses, therefore the reference pulses pass through a separate 4f-grating stretcher. It introduces the same amount of GVD as the first, but the selected bandwidth is slightly wider. Moreover, nonlinear effects in the fiber are avoided as well by using stretched pulses. Each amplified pulse and its corresponding reference pulse is selected out of the full FC pulse train with two Pockels cells (PC), which are used in double pass (with a combined contrast of 1:$10^4$) to reduce background light. A third PC is used in combination with polarization optics to introduce a slight vertical offset between each set of pulses in a small delay line. In this manner the interference pattern of the two sets of pulses can be projected above each other on a CCD camera (IMI-TECH IMB-716-G) so that both spectral interference patterns can be observed simultaneously. An example of the observed interferograms is shown in the lower panel of Fig.~\ref{fig:RC_laser_setup}. The spectrometer consists of a gold grating of 1200 lines/mm (Richardson, 53-*-360R) in combination with a 350 mm lens, which projects the spectrum on the camera. The period of the interference pattern can be adjusted by changing the delay between the original FC pulse and the amplified pulse in the delay line of the reference arm. The geometrical phase difference arising from slight misalignments of the two interference patterns on the camera is calibrated by exchanging the projection of the two pulses using the third PC. The phase difference between the pulses is extracted from the interferograms using a Fourier transform-based method \cite{takeda_fourier-transform_1982,kandula_ultrafast_2008}.	
	\subsection{The spectroscopy setup}
	\label{subsec:spectroscopy_setup}
	A schematic overview of the vacuum setup is shown in Fig.~\ref{fig:vacuum_setup}. High-harmonics are created in a gas jet of argon atoms in the first vacuum chamber. The jet is created by a supersonic expansion from a home-built piezo valve (type 1), which operates at a backing pressure of 5 bar. Before HHG, the center part of the amplified beam is cut-out with a 1-mm-diameter beam block at a distance of $2f$ before the focusing lens (with a focal distance $f = 250$ mm). This leads to a donut-shaped intensity profile of the beam which is then imaged in the second vacuum chamber to separate the harmonic light from the fundamental light with an adjustable iris. The generated harmonics propagate on axis with a much lower divergence than the fundamental. Therefore, the iris transmits the harmonic beam, while it blocks the ring of fundamental light around it.\\ 
	A peak-intensity of up to $ 1.3 \times 10^{14}$ W/cm$^{2}$ is reached in the interaction region below the nozzle of the valve. The position of the valve can be adjusted in all three dimensions to optimize the harmonic yield. In particular, the alignment in the direction of propagation is important for reaching proper phase-matching conditions of the HHG process. This is achieved by adjusting the valve position such that the divergence of the harmonic beam is minimized, which typically leads to a focus position just in front of the jet.\\
	Due to the low ionization potential of xenon, direct one-photon ionization is possible with harmonics of order $q>7$. Therefore, a 1-mm-thick lithium fluoride window is placed in the beam, which only transmits light at $\lambda>105$ nm. An absolute transmission efficiency of 40 $\%$ was measured for the seventh harmonic at 110 nm.\\
	A toroidal mirror pair operating at a grazing angle of 7.5$^\circ$  is used to refocus the harmonic beam in the interaction region. It forms a one-to-one telescope with an effective focal length of 250 mm. Coma and other aberrations are fully compensated at equal angle of incidence, but for a slight angular deviation the beam quality in the spectroscopy region is significantly perturbed. A retractable silver mirror is placed a few centimeters behind the second toroidal mirror to monitor changes in day-to-day alignment and ensure a proper quality of the refocused fundamental beam. The harmonic beam itself can be detected in the far field (400 mm from the focal plane) with an XUV sensitive CCD camera (Andor Newton SY, DY940P).\\
	The xenon atoms are excited at a 90$^{\circ}-$angle in a pulsed atomic beam to reduce influences from the first-order Doppler effect. The gas pulse is created from a home-built piezo valve (type 2, see \cite{trimia_short_2009}) with a 0.3 mm nozzle opening, operating at 0.5 bar backing pressure. The formation of clusters from the gas expansion, as reported in previous studies \cite{wormer_fluorescence_1989,ozawa_vuv_2013}, was thoroughly investigated by changing the parameters of the expansion (backing pressure, pulse timing, pulse length) and by changing skimmers and valve types (solenoid and piezo based with different valve openings from 0.15 mm to 0.8 mm), but no evidence for the production or influence of clusters was found in the Ramsey signals or time-of-flight (TOF) mass spectrometer data. After these tests we based the atomic beam geometry for the measurements on two skimmers to reduce Doppler broadening. The first skimmer is placed at a distance of 200 mm from the valve and has a circular aperture with a diameter of 8 mm. This relatively large opening was chosen to reduce possible effects of skimmer clogging. The second one is an adjustable slit skimmer set to a width of 1 mm. It is made from 0.150-mm-thick stainless steel foil and bent into a slit skimmer pointing towards the source. The slit is orientated perpendicular to the atomic and VUV beam. It is placed at a distance of 460 mm from the valve, leading to a maximum atomic beam divergence of $<1.5$ mrad.\\ 
	The excited xenon atoms are selectively ionized by a 3 mJ, 60 ps ionization pulse at 1064 nm, which is delayed by 2 ns with respect to the second excitation pulse. This pulse is derived from the second pump laser pulse, and therefore automatically shifts with the different pulse delays which are selected for the RC measurement. The ionization beam is focused in the vertical direction with a cylindrical lens to match the shape of the interaction region and to reach the saturation intensity of the ionization step. The ions are extracted through a 40 cm long TOF drift tube, separating the different isotopes temporally. They are measured by a fast electron-multiplier (ETP ion detect, AF880) and the signal from the selected isotope is obtained with a boxcar integrator (Stanford Research). The whole experiment is repeated at a rate of 28.2 Hz.
	
	\section{Results}
	\label{sec:results}
	\begin{figure}[!t]
		\centering
		\includegraphics{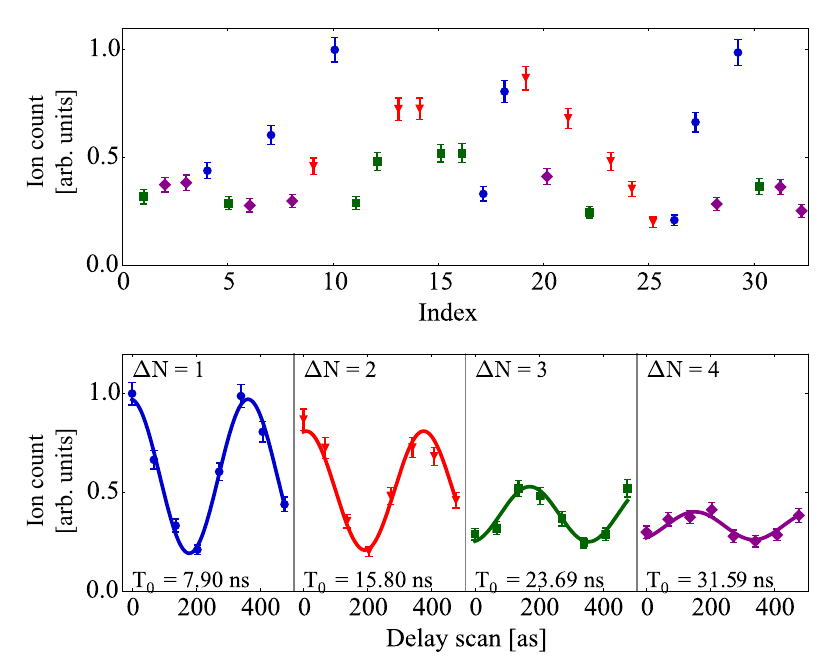}
		\caption{A typical Ramsey-comb scan at $\Delta N$=~1-4. The upper panel shows the signal as a function of measurement number, where the different markers show the data points corresponding to a specific $\Delta N$. In the lower panel, the data is sorted according to pulse delay, leading to clear Ramsey fringes. The signal was obtained by scanning the pulse delay over 475 attoseconds, while a 7.9 ns delay is introduced between each Ramsey measurement by changing the pulse pair. }
		\label{fig:RCS_example}
	\end{figure}
	A typical Ramsey-comb scan of the $5p^6 \rightarrow 5p^5 8s~^2[3/2]_1$ transition is shown in Fig.~\ref{fig:RCS_example}. The lower panel shows four Ramsey fringes at $\Delta N = 1-4$. These are obtained by measuring the number of ions, which is directly proportional to the number of excited atoms, as function of pulse delay. Each point is obtained by averaging over $\sim 500$ laser shots and the error bars represent the observed standard deviation of the mean value. The total scan range was 475 attoseconds in which 1.3 fringe was observed. The phase difference between the two FC pulses drifts on the order of 10~mrad on a ten minute timescale, due to subtle changes (mostly related to temperature) in the amplifier system. Therefore the measurement time was set to be only 3.3 minutes/fringe at the expense of signal-to-noise ratio, and only pairs of Ramsey fringes were used ($\Delta N = 2$ and $\Delta N = 4$). Additionally the data points were recorded in random order, as shown in the upper panel of Fig.~\ref{fig:RCS_example}, reducing the influence of drifts further by a factor of $\sim$4.\\
	
	\subsection{The Ramsey fringe contrast}
	\label{subsec:RCS_contrast}
	The contrast of the Ramsey fringes is influenced by several processes. An overall reduction of the Ramsey fringe contrast is caused by the phase noise on the amplified and up-converted pulses, because it is mostly produced by the parametric amplification process so that the noise amplitude can be considered constant for a large range of pulse delays. A delay-dependent ($\Delta N T_{rep}$) decay of the Ramsey fringe contrast is affected by: the finite upper state lifetime (22 ns~\cite{chan_absolute_1992}), Doppler broadening, a wavefront tilt of the excitation beam and the short transit time of the xenon atoms through the refocused excitation beam. The latter was the dominant cause for a relatively fast decay of the Ramsey fringe contrast in this experiment, because the one-to-one toroidal telescope refocuses the HHG beam to a spot similar in size as at the HHG source, which is $\leq 50~\mu$m for the fundamental beam and typically a few times smaller in the VUV. This configuration was chosen in preparation for $1S-2S$ excitation in a trapped He$^+$ ion (requiring refocusing), but was not ideal for the current xenon atomic beam experiment. Therefore the toroidal mirrors were deliberately slightly misaligned (with respect to perfect one-to-one imaging configuration) to introduce a small amount of astigmatism, which increases the transit time at the expense of a local wavefront tilt. As a result, the maximum pulse delay was increased from 16 ns (diameter of $\sim$15 $\mu$m) to 32 ns ($\sim$30 $\mu$m).
	\begin{figure}[!t]
		\centering
		\includegraphics{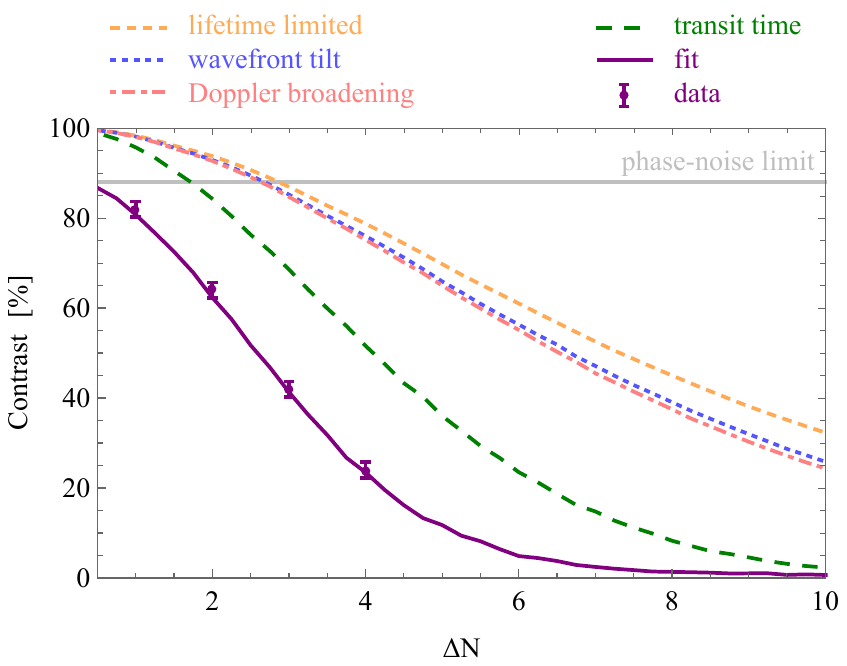}
		\caption{The measured Ramsey fringe contrast (purple points) as a function of pulse delay (equal to $\Delta N T_{rep}$) and the best fit (purple curve). The other curves show the influence on the contrast for each effect separately. These are obtained by assuming an upper state lifetime of 22 ns (dashed orange curve), a wavefront tilt ot 1.25 mrad (dotted blue curve), an atomic beam divergence of 1.5 mrad (dashed-dotted pink curve) and an effective interaction region of 26 $\mu$m (dashed green curve). The curves clearly show that for this experiment the dominant contrast-reduction effect was caused by the short transit time of the atoms through the interaction region. The fit also includes an additional phase noise amplitude of 400 mrad at the seventh harmonic, leading to a maximum contrast of 88\% (gray line).}
		\label{fig:contrast_delay}
	\end{figure}
	The purple data points in Fig.~\ref{fig:contrast_delay} show the measured Ramsey fringe contrast at different inter-pulse delays (with $\Delta N = 1-4$) after increasing the interaction region.\\ 
	In order to gain more insight into the influence of the individual contributions that affect the Ramsey fringe contrast, the Ramsey-comb signal was simulated and a comparison with the data was made. The simulations were performed with an upper state lifetime of 22 ns~\cite{chan_absolute_1992} and the maximum achievable contrast based on this value is shown with the finely dashed orange curve in Fig.~\ref{fig:contrast_delay}. Other parameters that were fixed based on prior knowledge were the maximum atomic beam divergence of 1.5 mrad (dashed-dotted pink curve) and the wavefront tilt of 1.25 mrad (dotted blue curve) based on the introduced astigmatism in the VUV beam. The latter two were estimated from the geometrical configuration. The remaining two parameters, namely the size of the interaction region and the phase noise were fitted to match the decay and the offset of the measured contrast. As discussed before, the phase noise manifests itself as an overall reduction of the Ramsey fringe contrast, while the limited transit time affects the decay of the curve. The purple line shows the best match with the experiment and leads to a 26 $\mu$m VUV beam size at full width half maximum and a 400 mrad phase noise amplitude at the seventh harmonic. From the latter an absolute maximum contrast of 88\% is expected (gray line). The obtained value indicates that the phase noise amplitude of the amplified NIR pulses is 57 mrad, taking the linear scaling with the harmonic order into account. This result agrees well with the extracted value from direct phase measurements at the fundamental frequency, which will be discussed in Sec.~\ref{subsec:ab_trans_freq}. Even though the beam size was increased by almost a factor of two, the loss of contrast is still dominated by the short transit time of the atoms (dashed green curve).\\ 
	Due to the high sensitivity to the transit time of the atoms in this experiment, it became apparent that the two amplified pulses initially had a slightly different propagation direction. This was caused by an intensity dependent spatial walk-off effect in the parametric amplifier.\\ 
	The two pump pulses of the NOPCPA have a slightly different intensity profile as a result of saturation effects in the amplifiers of the pump laser~\cite{morgenweg_multi-delay_2013}. This leads to a different walk-off induced propagation angle for the two amplified FC pulses caused by pump-to-signal phase transfer. This problem was solved by implementing a walk-off compensating configuration for the first two passes \cite{armstrong_parametric_1997}, which is based on the reversal of the walk-off direction between the first two crystals by rotating one of the crystals around the proper axis. In this manner the beam pointing difference was reduced from up to $\sim 0.5$ mrad (depending on the day-to-day alignment) to 20 $\mu$rad. Note that previous RCS measurements~\cite{altmann_deep-ultraviolet_2018,altmann_high-precision_2016} were not affected by this because they used large collimated beams in combination with \emph{low} harmonic generation.
	\subsection{Plasma-induced effects}
	\label{subsec:plasma_induced_effects}
	\begin{figure}[!t]
		\centering
		\includegraphics{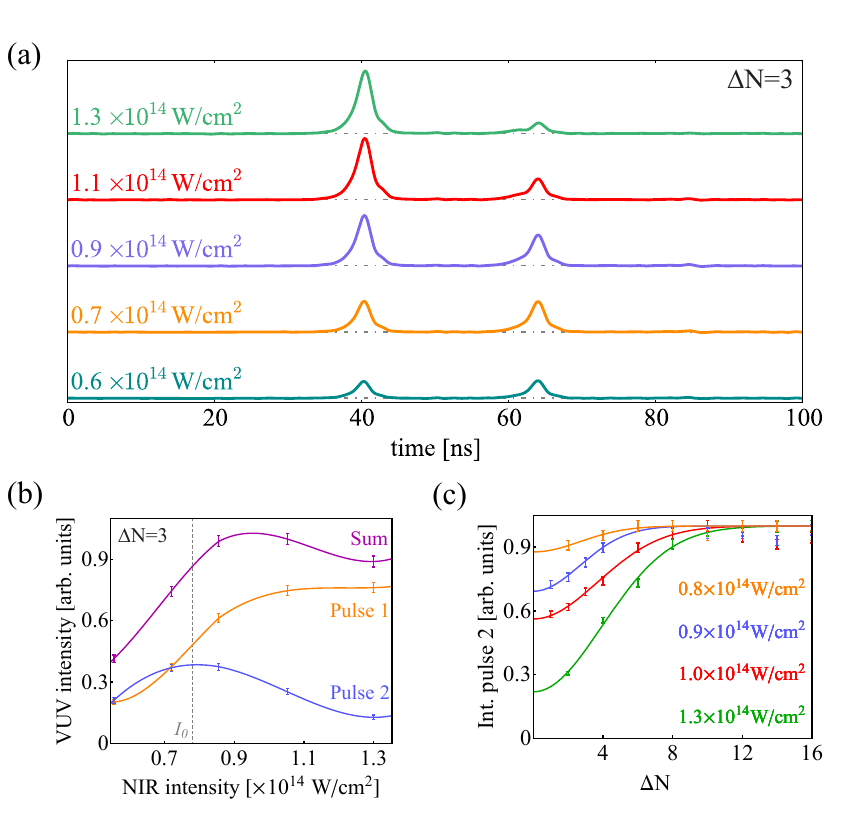}
		\caption{(a) The recorded seventh-harmonic VUV yield of the two individual pulses at $\Delta N = 3$ for different driving intensities. Initially both pulses increase as a function of driving intensity, but above $0.9 \times 10^{14}$W/cm$^2$ the yield of the second pulse stagnates and eventually decreases due to the depletion of the neutral atom density. (b) The amplitude of the individual VUV pulses at $\Delta N = 3$ and sum of the two. The lines connecting the data points are splines to guide the eye. (c) The extracted amplitude of the second VUV pulse as a function of inter-pulse delay (equal to $\Delta N T_{rep}$) for different driving intensities. The lines are fits to the data assuming the fundamental pulses have a Gaussian beam profile. A revival of the second pulse occurs on a timescale of 50-100 ns.}
		\label{fig:plasma_intensity_effect}
	\end{figure}
\begin{figure}[!t]
		\centering
		\includegraphics{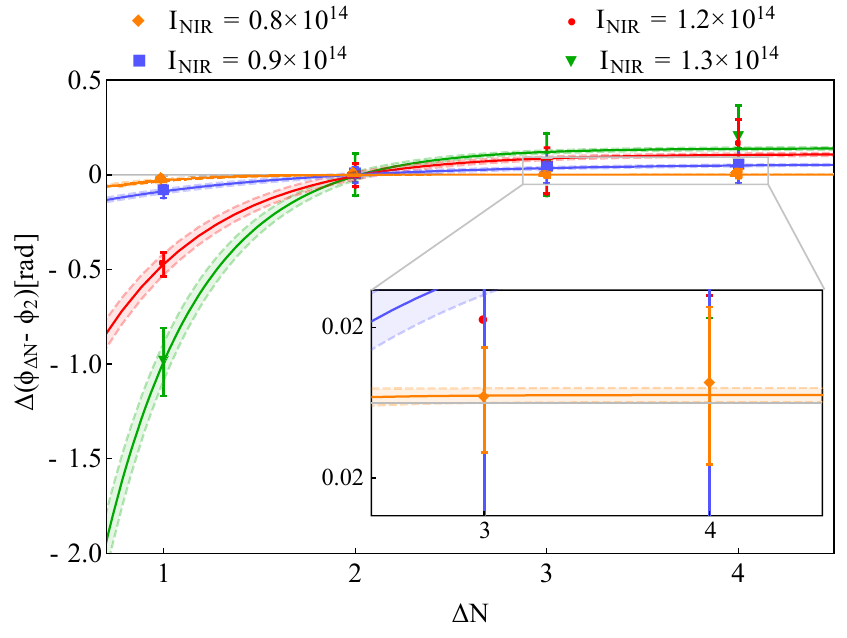}
		\caption{The observed phase shift at $\Delta N = 1,3$ and $4$ relative to $\Delta N = 2$ as a function inter pulse delay and for different driving intensity. The data is fitted with $\Delta \phi \propto e^{-B \Delta t}$ and the shaded area represents the $1\sigma$ uncertainty interval. The offset in the zoom-in shows the residual shift at large delay for normal operating conditions ($I_0 = 0.78 \times 10^{14}$).}
		\label{fig:plasma_induced_phase_shift}
	\end{figure}

	The influence from plasma formation on the HHG process was investigated by varying the driving intensity with a half-wave plate and a thin-film polarizer. As illustrated in Fig.~\ref{fig:plasma_intensity_effect}, the depletion of the neutral atoms after up-conversion of the first pulse influences the VUV yield of the second pulse. Fig.~\ref{fig:plasma_intensity_effect}(a) shows the individual seventh-harmonic VUV pulses at $\Delta N =3$ for five different driving intensities. Up to $0.8\times 10^{14}$W/cm$^2$, the yield of both VUV pulses benefit from a higher driving intensity, but a further increase leads to the stagnation and eventually a decrease of the yield of the second pulse. At the highest driving intensity ($1.3 \times 10^{14}$W/cm$^2$) the second VUV pulse is almost completely suppressed. Note that in this regime, the phase matching condition, which depends on the dispersion, is also significantly influenced. As a result also the total VUV yield, i.e.~the sum of the two pulses at $\Delta N = 3$, stagnates as is shown in Fig.~\ref{fig:plasma_intensity_effect}(b). Above $0.9 \times 10^{14}$W/cm$^2$ there is no significant improvement of the total yield.\\
	A revival of the second VUV pulse occurs for larger inter-pulse delays, as shown in Fig.~\ref{fig:plasma_intensity_effect}(c), because the plasma moves out of the interaction region and the gas sample refreshes as is also indicated in Fig.~\ref{fig:vacuum_setup}. The typically observed timescale on which this revival occurs, depends on the size of the interaction region and therefore the driving intensity. The reason for this is that a higher driving intensity leads to a bigger volume where the intensity is high enough for tunnel ionization and therefore HHG. It then also takes longer for all the ions to leave the interaction zone. Recovery times of 50-100 ns were observed as is shown in Fig.~\ref{fig:plasma_intensity_effect}(c), which matches well with what is expected for a focus diameter in the HHG zone of about 50 $\mu$m and the estimated velocity of the argon atoms from a supersonic expansion ($v=550$ m/s).\\
	The delay-dependent influence from plasma formation on the phase of the second VUV pulse was extracted from the phase evolution of the Ramsey fringes as function of delay (which is schematically depicted in the top panel of Fig~\ref{fig:vacuum_setup}) for different driving intensities. The absolute phase of the VUV Ramsey fringes drifts over longer timescales (typically a few mrad per minute) and therefore only the relative phase between the fringes was used such that this effect did not appreciably influence the measurements. The phase at $\Delta N = 2$ was chosen as a reference, because a remarkable change in the dynamics was observed after this delay (16 ns). Furthermore, the signal-to-noise is better at $\Delta N = 2$ than at longer delays and therefore subtle shifts could be measured more accurately in this manner.\\ 
	Fig.~\ref{fig:plasma_induced_phase_shift} shows the measured phase shift as a function of delay where $\Delta N = 2$ has been taken as a reference value. The curves are fitted with an exponential function $\Delta \phi \propto e^{-B \Delta t}$.
	At the shortest delay ($\Delta N =1$), we observe a noteworthy seventh-order dependency of the phase-shift as function of peak intensity (for more details, see \cite{dreissen_high-precision_2019}), leading to a shift of 1 rad at the highest deployed peak intensity. This shift is significantly reduced for larger inter-pulse delays, indicating that there are much faster dynamics involved in this process than what is expected from the dynamics of ions and atoms (Fig.~\ref{fig:plasma_intensity_effect}(c)). Therefore we conclude that the phase effects are predominantly caused by free electrons in the plasma, which can leave the interaction zone much faster than the atoms and ions (on a picosecond-timescale instead of a nanosecond-timescale). This is in agreement with the expected dominant contribution to the change in refractive index due to plasma formation~\cite{popmintchev_phase_2009,rundquist_phase-matched_1998}. Note that the error bars increase for higher driving intensity because the two excitation pulses (Fig.~\ref{fig:plasma_intensity_effect}(a)) and therefore the excitation contributions become unequal, which directly leads to a lower contrast of the Ramsey fringes.\\
	The asymptotic value at $\Delta N \rightarrow \infty$ represents the residual phase shift at $\Delta N = 2$, which is significant at high driving intensity (140(74) mrad at $1.3\times 10^{14}$~W/cm$^2$), but reduces to a negligible level of 2(2) mrad at the lowest driving intensity. This is more clearly visible in the inset of Fig.~\ref{fig:plasma_induced_phase_shift} which shows a zoom-in of the induced phase shift for low driving intensity at large pulse delay. By skipping $\Delta N =1$ and moderating the intensity to $0.78\times 10^{14}$~W/cm$^2$, the observed plasma induced phase shift could be reduced to -2(5) mrad between $\Delta N = 2$ and $\Delta N = 3$, and -7(9) mrad between $\Delta N = 2$ and $\Delta N = 4$. The corresponding frequency shift for the measured transition of -32(91) kHz and -67(86) kHz, respectively, is consistent with zero within the uncertainty. Fig.~\ref{fig:plasma_intensity_effect}(b) shows that the moderation of the intensity to $I_0$ leads to a reduction of 15$\%$ of the total VUV yield and therefore only a slight reduction of the total signal level. The VUV peak intensity at the seventh harmonic was estimated to be $5 \times 10^7$ W/cm$^2$ at these typical operating conditions for spectroscopy.
	\begin{figure}[!t]
		\centering
		\includegraphics{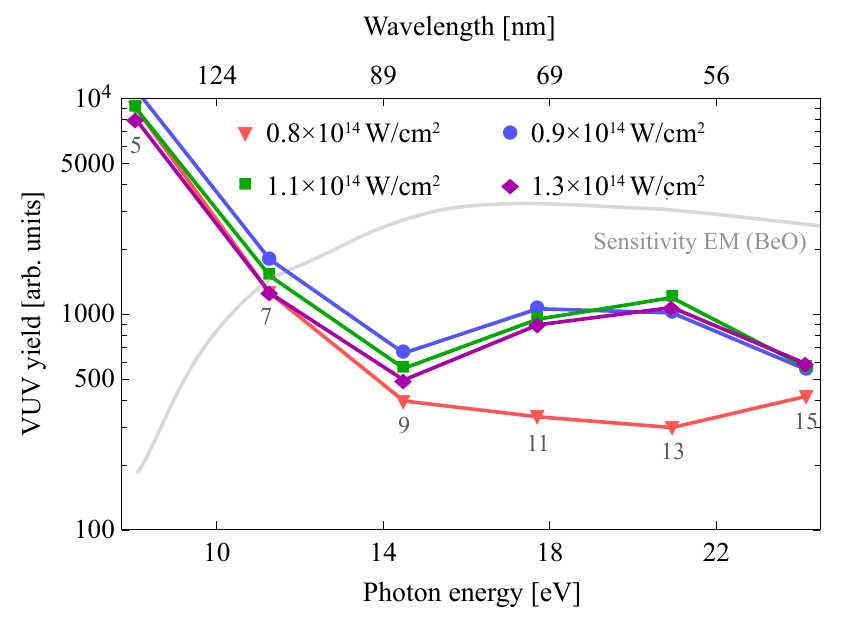}
		\caption{The obtained harmonics spectrum between order $q=5$ (154 nm) and $q=15$ (51 nm) for four different values of the driving intensity. The individual points are the measured intensities at the indicated (by the number below the points) harmonic order. The curves are corrected for the sensitivity of the Electron Multiplier (EM) consisting of BeO (gray curve), but not for the unknown wavelength-dependent efficiency of the grating, so it only gives an indication of the harmonic yield. Generation of the highest harmonics $q>7$ indicate that the region of tunneling ionization is reached, even for low driving intensity. The different power scaling behavior of the 11$^{\mathrm{th}}$ and 13$^{\mathrm{th}}$ harmonic is reproducible, but not fully understood.}
		\label{fig:VUV_spectrum}
	\end{figure}\\
 	The spectrum of the generated harmonics at the different values of the driving intensity was measured with the monochromator at the end of the vacuum system (Fig.~\ref{fig:vacuum_setup}). The spectra are acquired with only a single pulse and do not represent the total yield of the double-pulse configuration. Fig.~\ref{fig:VUV_spectrum} shows four spectra obtained at different driving intensities. The harmonics of orders between $q=5$ (154 nm) and $q=15$ (51 nm) are clearly observed. The high-frequency end of the spectrum is limited by the poor reflectivity of the aluminum grating that was used (model 33009FL01-510H from Newport) at a relatively small angle of incidence ($<45^{\circ}$). The geometry of the monochromator prevents the low-frequency end of the spectrum to be observed, therefore the third harmonic is missing. The measured amplitudes of the harmonic orders were corrected for the wavelength dependent sensitivity of the electron multiplier (gray curve in Fig.~\ref{fig:VUV_spectrum}). However, the diffraction efficiency ($-1^{st}$ order) of the grating is unknown, because the grating is only specified down to 130 nm. Therefore the curves in Fig.~\ref{fig:VUV_spectrum} only give an indication of the spectral intensity of the different harmonic orders. Nevertheless two conclusions can be drawn from the observed spectra. The first is that the highest harmonics are produced even at the lowest driving intensity, indicating that the regime of tunnel ionization is reached in the HHG process at those levels of intensity. Secondly, there is an increase of about a factor of 4 in harmonic yield for wavelengths below 80 nm at higher intensities. For future experiments (like $1S-2S$ He$^+$ excitation) where the pulses will be more than 100 ns apart, one can fully use this improved HHG yield because in that case there is no influence of atom depletion or related phase shift on the second pulse. It is not clear why there is a (reproducible) sudden change in power scaling behavior for the 11$^{\mathrm{th}}$ and 13$^{\mathrm{th}}$ harmonic. It could in part be induced by changing phase matching conditions (optimized for the seventh harmonic) due to ionization and perhaps subtle pulse shape changes (induced by the optics used to vary the intensity).
 	
	\subsection{Calibration of the absolute transition frequency}
	\label{subsec:ab_trans_freq}
	\begin{figure}[!t]
		\centering
		\includegraphics{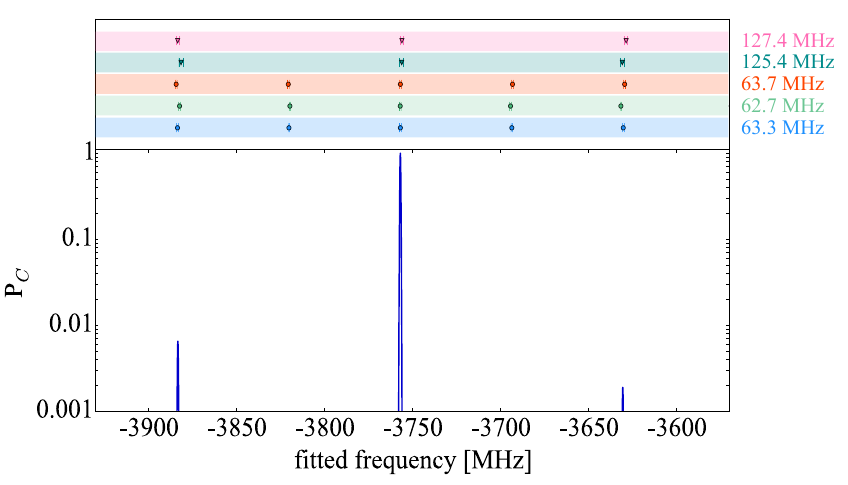}
		\caption{The upper panel shows the extracted possible transition frequency with respect to the previous value of $2\,726\,090(5)$ GHz~\cite{yoshino_absorption_1985} at 5 different effective mode spacings, which are indicated in the upper-right corner. The lower panel shows the probability of the coincidence point at the possible frequencies. The coincidence point with the highest probability (normalized to 1) is indicating the true transition frequency, and was determined with 99.2$\%$ confidence.}
		\label{fig:coincidence_point}
	\end{figure}
	The measured plasma-induced phase shifts indicate that high-precision spectroscopy is possible, even for RCS with relatively small pulse delays. To demonstrate this and show the full potential of RCS in combination with HHG, the absolute transition frequency of the $5p^6 \rightarrow 5p^5 8s~^2[3/2]_1$ transition has been determined.\\ 
	Due to the periodicity of the signal in Ramsey-comb spectroscopy, the extracted transition frequency is only known modulo the effective mode spacing ($f_{rep}/\Delta N$). Therefore, the repetition frequency was varied over a range of 1.6$\%$ in three steps to determine the actual transition frequency. A few of the possible values for the transition frequency are shown in Fig.~\ref{fig:coincidence_point} relative to the previous determination from Yoshino \textit{et al.} \cite{yoshino_absorption_1985}. The upper panel shows the extracted frequencies at five different values of the mode spacing. The results obtained from the combination of $\Delta N = 2$ and $\Delta N= 4$ yields the best statistical accuracy but reduces the effective mode spacing to $f_{rep}/2$, which leads to twice as many possible transition frequencies. Therefore, half of the possibilities were excluded by also performing measurements at $\Delta N = 2$ and $\Delta N = 3$. The combination leads to only a single coincidence point of the transition frequency within a 4$\sigma$ range of the previous determination \cite{yoshino_absorption_1985}.\\
	The probability that the obtained transition frequency is at the given coincidence point is calculated by constructing Gaussian distributions from the extracted transition frequencies ($\mu_{n}$) and the corresponding uncertainties ($\sigma_{n}$) according to
	\begin{equation}
	P_n(f) = \frac{1}{\sigma_n \sqrt{2 \pi}}e^{-\frac{(f-\mu_n)^2}{2\sigma_n^2}}. 
	\end{equation}
	The distributions obtained with different mode spacing are multiplied to obtain the probability ($P_C$) that the transition frequency is at a certain coincidence point, which is shown in the lower panel of Fig.~\ref{fig:coincidence_point}. In this manner, the obtained coincidence point was determined with 99.2$\%$ confidence.\\
	After identification of the proper transition frequency (`mode'), the absolute transition frequency was determined by evaluating a series of systematic effects. The main systematic effect is caused by the first-order Doppler shift. This was calibrated by comparing the transition frequency obtained from a beam of pure xenon with that obtained from a mixture of argon and xenon in a ratio of 3 to 1. The mixture accelerates the xenon atoms, which unfortunately also reduces the transit time so that only a maximum pulse delay of $\Delta N=3$ could be used in this case.\\
	In the TOF mass spectrometer geometry, two deflection plates were included in the direction of propagation of the atoms to compensate for the forward velocity and to steer the ions onto the detector. The observed signal level as a function of deflection voltage was compared to simulated trajectories (and resulting  signal) using COMSOL to calibrate the forward velocity. From this we established a speed of 285(30) m/s for pure xenon, and 480(30) m/s for xenon in the mixture.
	\begin{figure}[!t]
		\centering
		\includegraphics{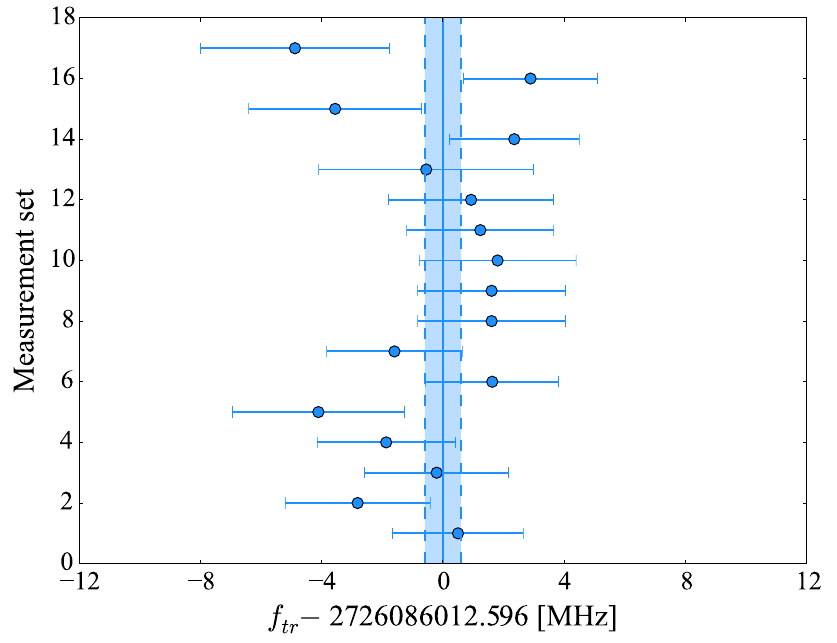}
		\caption{The obtained Doppler-free transition frequency. Each data point shows the extracted transition frequency from 10 measurements with pure xenon and 10 measurements in the 3:1 Ar:Xe mixture. The shaded area shows the 1$\sigma$ uncertainty interval.}
		\label{fig:doppler_free_frequency}
	\end{figure}\\ 
	The atomic beam was first coarsely aligned to be perpendicular to the excitation beam by minimizing the observed frequency shift to the level of a few MHz. The Doppler-free transition frequency was then obtained by extrapolating the residual shift to zero velocity after correction for the HHG shift and the second-order Doppler shift (1.2 kHz for pure xenon and 3 kHz for the mixture). In total more than 300 measurements were recorded over a period of 17 days from which the final result was obtained, which is shown in Fig.~\ref{fig:doppler_free_frequency}.\\
	As was discussed in Sec.~\ref{subsec:RCS}, the recoil from absorption of a single photon affects the extracted transition frequency. Therefore, a correction of 125 kHz was applied in the evaluation of the final result.
	\begin{figure}[!t]
		\centering
		\includegraphics{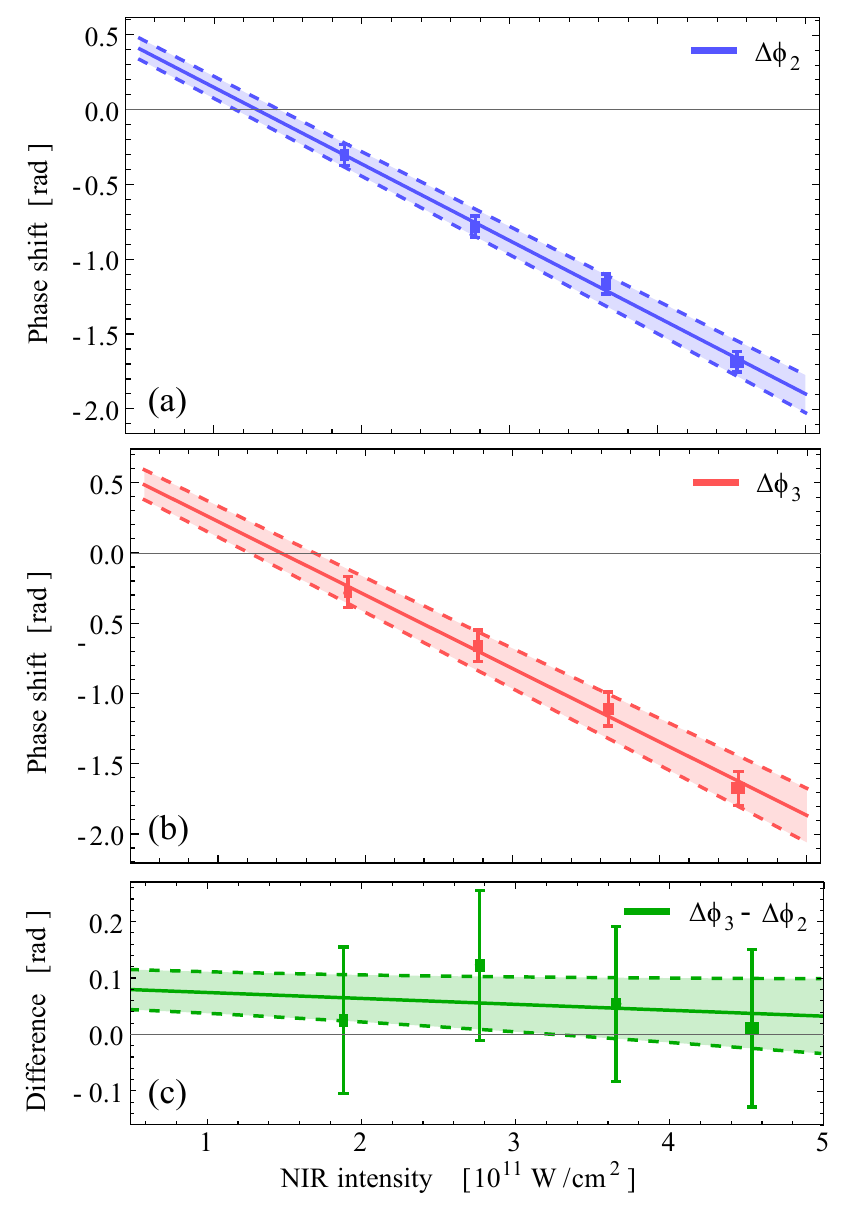}
		\caption{The observed phase shift of the Ramsey fringe in the VUV as a function of NIR intensity in the interaction region for (a) $\Delta N = 2$, (b) $\Delta N = 3$ and (c) their difference. The solid line represents a linear fit of the data and the shaded area indicates the $1\sigma$ uncertainty interval. A maximum shift of 1.65 rad was observed by increasing the intensity in the interaction region by a factor of 4.6. However, the fitted slopes of the curves at $\Delta N = 2$ and $\Delta N = 3$ are almost equal, with a small residual difference between $\Delta N=2$ and $\Delta N =3$, as is shown in (c), due to drifts of the setup as the measurements where performed sequentially (see text). }
		\label{fig:AC_stark}
	\end{figure}\\
	Although the ac-Stark shift is known to be significantly suppressed in the RCS method \cite{morgenweg_ramsey-comb_theory_2014,altmann_high-precision_2016}, we have investigated the presence of possible residual shifts. Both the VUV pulses and the residual light from the fundamental beam can induce a shift of the energy levels in the atom. To determine the magnitude of the effect, the induced phase shift of a single Ramsey fringe was measured as a function of intensity.\\ 
	The influence of the residual NIR light was measured by slightly opening the iris (in the second vacuum chamber) that normally blocks as much of the NIR beam as possible. For normal operating conditions the NIR intensity in the interaction region was estimated to be $1 \times 10^{11}$ W/cm$^2$. The measured phase shift as function of NIR intensity for $\Delta N = 2$ and $\Delta N = 3$ is shown in Fig.~\ref{fig:AC_stark}(a) and (b), respectively, and the solid line shows a linear fit to the data of which the $1\sigma$ uncertainty interval is indicated by the shaded area. A significant intensity-induced phase shift of 1.65 rad was observed by increasing the intensity by a factor of 4.6. However, the differential nature of RCS leads to a strong suppression of the influence of this effect on the extracted transition frequency. This is shown Fig.~\ref{fig:AC_stark}(c), where the difference between the induced shift at $\Delta N = 2$ and $\Delta N =3$ is plotted. The fit shows that the dependency of the phase on the intensity is practically equal (within the statistical uncertainty) for both pulse delays (a small difference is seen because the measurements were performed sequentially instead of simultaneously and without random scanning, which leads to a small residual phase drift). Therefore the effect is common mode and barely affects the extracted transition frequency. Given that the NIR pulses are actively kept constant to the level of 0.2$\%$, the ac-Stark effect is suppressed by a factor of 500, leading to a frequency shift of less than $20$ kHz.\\
	A similar procedure was performed to calibrate the ac-Stark shift induced by the VUV pulses. Here the intensity was decreased with a factor of three compared to normal operating conditions (estimated at $5\times10^7$ W/cm$^2$ for the seventh harmonic) by reducing the density in the HHG jet to reduce the third to seventh harmonic output roughly equally. To this end the gas pressure in the HHG interaction zone was lowered by reducing the drive voltage on the piezo valve. A phase shift of 0.7 rad was observed, which, in combination with the measured stability of the VUV intensity (about ten times worse than that of the NIR), results in a frequency uncertainty of $85$ kHz.\\
	The dc-Stark effect was suppressed by using a pulsed electric field to extract the xenon ions. A possible residual shift from stray electric fields was estimated by deliberately increasing the field in the interaction zone to 29 V/cm. The observed frequency shift of 2.5(1.0) MHz, gives rise to a shift of $< 20$ kHz assuming an estimated maximum stray field of $< \pm$ 0.25 V/cm.\\
	A similar procedure was performed to calibrate the Zeeman shift. A set of six coils was used to reduce the magnetic field in the interaction zone to below 0.1 Gauss. Based on measurements at 3 Gauss (in different directions), we determined the uncertainty from stray magnetic fields to be below $52$ kHz.\\
	Finally, the phase stability of the amplified FC pulses was measured using the setup described in Sec.~\ref{subsec:phase_measurement_setup}. The absolute phase shift introduced in the amplification process is typically a few hundred mrad. The observed phase noise amplitude was 40-70 mrad, depending on the exact operating conditions in the NOPCPA. This is in agreement with the estimation, based on the contrast of the Ramsey fringes, as was discussed in Sec.~\ref{subsec:plasma_induced_effects}. The origin of the phase noise was traced back to the amplification process itself \cite{ross_analysis_2002}. A clear correlation was observed between the output power of the amplified FC beam and the phase stability of the pulses, indicating that local wavefront fluctuations (and therefore phase mismatch) give rise to added phase noise, rather than effects such as self and cross phase modulation \cite{ross_analysis_2002,morgenweg_multi-delay_2013}.
	\begin{figure}[!t]
		\centering
		\includegraphics{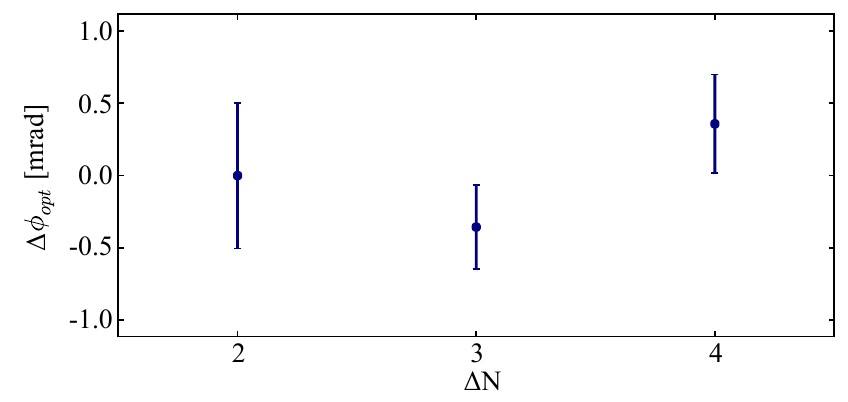}
		\caption{The relative optical phase shift of the fundamental beam at 770 nm as a function of the pulse delay (equal to $\Delta N T_{rep}$). The data points are obtained by averaging over 15 measurements, each with a statistical uncertainty of 1.5-2.5 mrad. This is equivalent to a typical measurement day. The average differential phase stability is considerably better than 1 mrad.}
		\label{fig:phase_shift_measurement}
	\end{figure}\\ 
	The stability of the phase as a function of pulse delay was measured as well, as this can have an influence on the extracted transition frequency. A set of 15 measurements, each with a statistical uncertainty of 1.5-2.5 mrad, were averaged to obtain the phase stability over a typical measurement day. The result is shown in Fig.~\ref{fig:phase_shift_measurement}. The phase difference remains constant well within 1 mrad as a function of the delay, which conservatively leads to an uncertainty of $140$ kHz on the transition frequency in the VUV.\\
	The final result, and all the contributions to the error budget, are shown in Tab.~\ref{tab:error_budget_xe}. The transition frequency of the $5p^6 \rightarrow 5p^5 8s~^2[3/2]_1$ transition in $^{132}$Xe was determined to be $2\,726\,086\,012\,471(630)$ kHz, which improves upon the previous determination by a factor of $10^4$ \cite{yoshino_absorption_1985}. The achieved fractional accuracy of $2.3 \times 10^{-10}$ is a factor of 3.6 better than the previous best spectroscopic measurement using HHG \cite{cingoz_direct_2012}.
	\begin{table}[h]
		\caption{Contributions (in kHz) to the measurement of the $5p^6\rightarrow5p^58s~^2[3/2]_1$ transition frequency in $^{132}$Xe.}
		\label{tab:error_budget_xe}
		\begin{ruledtabular}
			\begin{tabular}{lrl}
				&Value or correction	&(1$\sigma$)	\\ 
				\hline 
				Doppler-free transition frequency	&$2\:726\:086\:012\:596$	&(600)\footnote{Including the uncertainty of $\approx$ 90 kHz due to the residual HHG phase shift (see text) and the correction for the second-order Doppler shift of 1.2 kHz for pure Xe and 3.5 kHz for the mixture.}			\\
				Light induced effects				&0				&(87)			\\
				dc-Stark shift						&0				&(20)			\\
				Zeeman-shift						&0				&(52)			\\
				Amplifier phase shift				&0				&(140)			\\ 
				Recoil shift						&-125			&($10^{-7}$)			\\  
				\hline 
				Total 								&$2\:726\:086\:012\:471$	&(630)			\\
			\end{tabular}
		\end{ruledtabular}
	\end{table}\\
	\subsection{Isotope shift measurement}
	\label{sec:isotope_shift}
\begin{figure}[!t]
		\centering
		\includegraphics{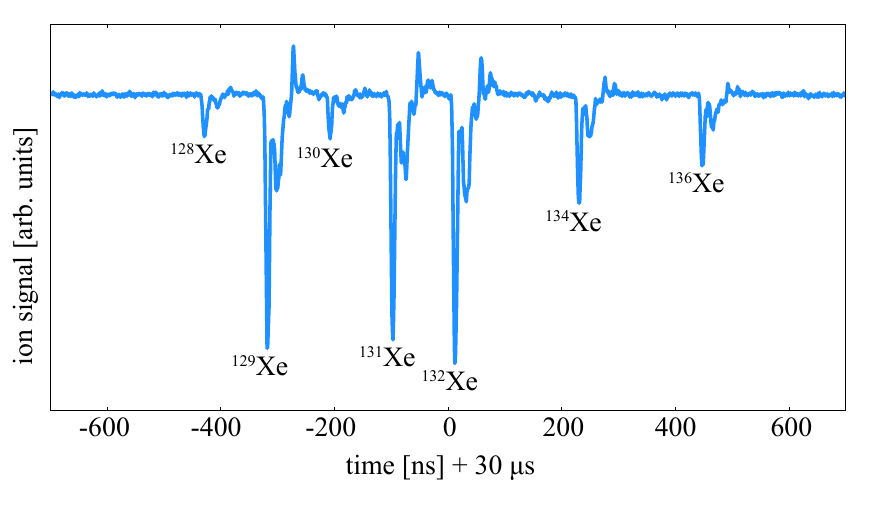}
		\caption{A typical time-of-flight (TOF) trace of xenon isotopes (using only one 110 nm pulse to avoid isotope-dependent interference effects). The relative amplitude shows the natural abundance of the isotopes. All stable isotopes, except for $^{124}$Xe and $^{126}$Xe, are clearly observed and resolved. Due to the high resolution of the TOF mass spectrometer, a double-peak structure was observed for all isotopes of which the origin could not be determined with certainty. It is most likely caused by different velocity classes in the atomic beam. To record Ramsey signals, only the first peak of each isotope was recorded with a Boxcar integrator in the experiment. }
		\label{fig:TOF}
	\end{figure}

\begin{figure}[!t]
		\centering
		\includegraphics{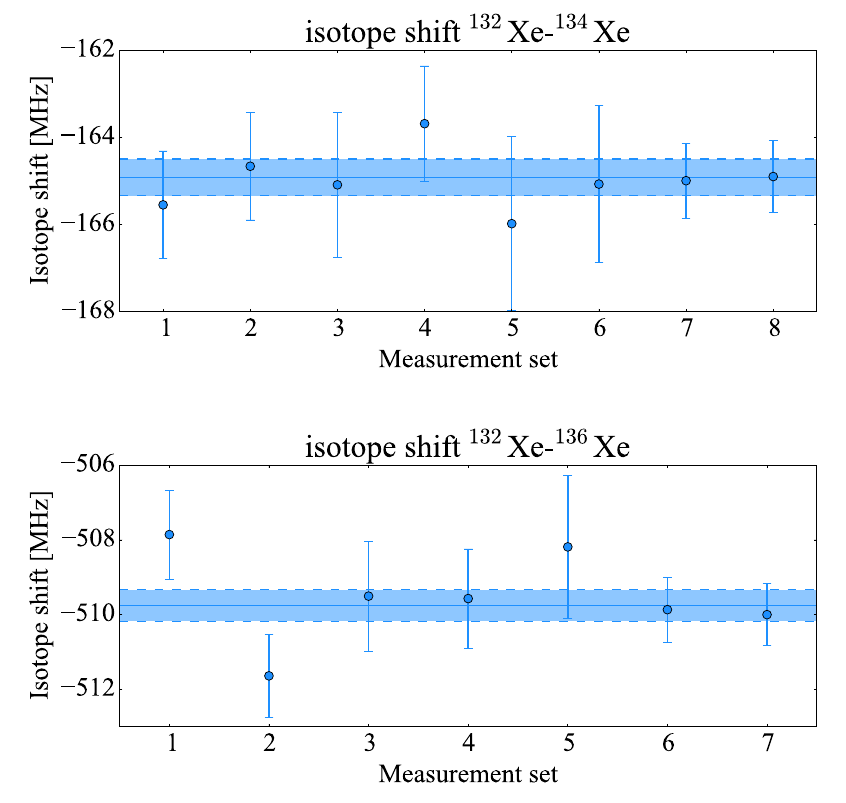}
		\caption{The measured isotope shift of the $5p^6 \rightarrow 5p^5 8s~^2[3/2]_1$ transition for $^{134}$Xe (upper panel) and $^{136}$Xe (lower panel), with respect to $^{132}$Xe. Each point is obtained from 8 simultaneous measurements of $^{132}$Xe and one of the other isotopes. The shaded area represents the $1\sigma$ uncertainty interval of the mean of the values. }
		\label{fig:isotope_shift}
	\end{figure}

	Xenon has seven observable stable isotopes, as shown in Fig.~\ref{fig:TOF}. Two of them have a nuclear spin and therefore hyperfine structure, namely $^{129}$Xe and $^{131}$Xe. This leads to excitation of several transitions simultaneously and therefore a beating of the corresponding Ramsey signals. In principle RCS can be used to obtain all the transition frequencies and the relative amplitudes of these contributions if the phase evolution can be measured over a sufficiently long delay \cite{morgenweg_ramsey-comb_2014,morgenweg_ramsey-comb_theory_2014}, but due to the short transit time of the xenon atoms this could not be realized, and therefore only the even isotopes were considered.\\ 
	Of the remaining five even isotopes, three provided enough signal to perform accurate measurements. Therefore the isotope shift of the $5p^6 \rightarrow 5p^5 8s~^2[3/2]_1$ transition for both $^{134}$Xe and $^{136}$Xe relative to isotope $^{132}$Xe was determined.
	\\
	As shown in Fig.~\ref{fig:TOF}, the signal from the different isotopes can be resolved easily and acquired separately using several Boxcar integrators. In this manner, several systematic shifts are common-mode for both isotopes and the isotope shift $f_{X}-f_{132}$ can straightforwardly be extracted. Note that each isotope clearly shows a double-peak structure, which only became apparent because of the high resolution of the mass-spectrometer (about 5 ns). The origin of this structure was investigated thoroughly by exchanging e.g.~the valve and the skimmers, but no conclusive reason for this observation was found. However, the most probable cause of the multi-peak structure is the existence of different velocity classes. Therefore, only signal coming from the first peak was acquired for each isotope with the Boxcar integrators.\\  
	The ambiguity of the determined frequency shift due to the effective mode spacing is again removed by varying the repetition frequency of the laser, similar as discussed in Sec.~\ref{subsec:ab_trans_freq}. This results in a single coincidence point for each isotope shift frequency which were both determined with $>$95$\%$ confidence. The isotope shifts have been obtained from more than 50 measurements for each isotope, which were taken over a period of 8 days. The final result of the extracted shift for both isotopes is shown in Fig.~\ref{fig:isotope_shift}. The isotope shift was determined to be $-164\,910(420)$ kHz and $-509\,750(425)$ kHz for $^{134}$Xe and $^{136}$Xe, respectively, relative to $^{132}$Xe.

	\section{Conclusion and Outlook}
	\label{sec:disc}
	We have given a detailed account of the first demonstration of Ramsey-comb spectroscopy in combination with HHG. 
	A key aspect is the accurate characterization of the time-dependent plasma-induced phase effects from HHG. This was investigated by exciting xenon atoms to a superposition state with two up-converted phase-locked laser pulses, while varying the conditions in the HHG process. The phase effects derived from the Ramsey-comb signal could be tracked with mrad-level accuracy and on nanosecond timescales. 
	It shows that the effects from ionization of the HHG medium (argon) on the phase is dominated at short timescales (of a few ns) by the influence of free electrons, leading to phase shifts of up to 1 rad at 8 ns pulse delay. However, we found that this effect can easily be reduced to a negligible level by increasing the pulse delay to at least 16 ns and moderating the driving intensity. This enabled us to measure the absolute transition frequency of the $5p^6\rightarrow5p^58s~^2[3/2]_1$ transition in $^{132}$Xe with RCS at 110 nm with sub-MHz accuracy. \\
	The final result of the transition frequency is $2\,726\,086\,012\,471(630)$ kHz. This is in agreement with the previous determination \cite{yoshino_absorption_1985}, but with a $10^4$ improved accuracy. The achieved relative accuracy of $2.3 \times 10^{-10}$ is unprecedented using HHG for up-conversion and improves upon the previous best measurement involving HHG by a factor of 3.6~\cite{cingoz_direct_2012}. \\
	This demonstration shows the potential of RCS in the VUV and XUV spectral range. Combining the accurate pulse control that the FC laser offers with the advantages of high-power amplified pulses results in straightforward single-pass up-conversion, a high excitation probability due to the high pulse energy, easy wavelength tunability, and enables detection methods, such as state-selective ionization, that can be made nearly background free and with high efficiency.\\
	The achieved accuracy of the presented experiment is almost completely limited by the short transit time of the xenon atoms through the excitation beam, because the setup was in fact designed for $1S-2S$ excitation in He$^+$ with a refocused XUV beam. The interaction time can therefore be increased by using a collimated excitation beam, which would immediately give rise to a higher accuracy as the largest contributions in the error budget scale down with increasing pulse delay. Alternatively,  the transit time can be increased by using a trapped atom (ion) such as He$^+$, which has the added benefit that Doppler effects can be reduced significantly. Therefore, this method shows great promise for $1S-2S$ spectroscopy of singly-ionized helium. The much higher harmonic order required for He$^+$ excitation (effectively the 26$^{\mathrm{th}}$ in a scheme combining 32 nm with 790 nm)~\cite{krauth_paving_2019} sets a more stringent limit on the allowed phase noise of the amplified FC pulses. However, the current system can produce pulse pairs with a relative phase noise of about 50 mrad rms, which would still lead to Ramsey fringes with contrast of $35\%$. Both the FC and parametric amplifier can be improved to reduce phase noise further, leading to higher contrast. Moreover, the upper-state lifetime of the $2S$ state is 1.9 ms, so that, compared to the present experiment, a much larger pulse separation can be applied.  Therefore there are good prospects for RCS to reach kHz-level accuracy in the VUV and XUV spectral range. This is of great interest for new high-precision tests of quantum electrodynamics theory, determination of fundamental constants such as the Rydberg constant or the alpha particle radius, and searches for physics beyond the standard model.\\
	
	\begin{acknowledgments}
		We thank Professor Dr. Fr\'{e}d\'{e}ric Merkt for valuable discussions during this experiment, and Rob Kortekaas and the Precision Mechanics and Electronic Engineering groups at the Faculty of Science of the Vrije Universiteit for their technical support.\\ 
		K. E. acknowledges the European Research Council for an ERC-Advanced grant under the European Union\textquotesingle s Horizon 2020 research and innovation programme (Grant Agreement No. 695677) and FOM/NWO for a Program Grant (16MYSTP).
	\end{acknowledgments}

	\bibliography{references}
\end{document}